\documentclass[a4paper,11pt]{article}
\usepackage{amsmath, amssymb, amsfonts,indentfirst,graphicx,color,anysize}
\marginsize{3cm}{3cm}{2cm}{2cm}
\title{
	\includegraphics[width=0.35\textwidth]{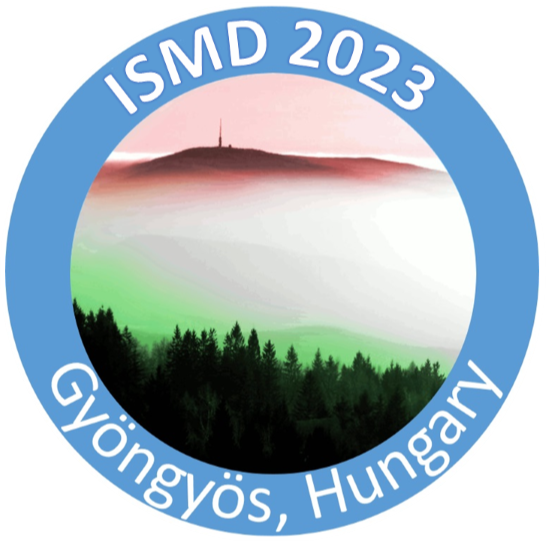}\\[1cm]
	\textbf{Relativistic two-particle problem in the Lagrange formalism }}
\author{{A.V.Koshelkin$^{1}$ }\\[1ex]
	$^1$National Research Nuclear University -MEPhI, Moscow, Russia\\
	}
	
	\usepackage{color}   %option added by cyw eg \pagecolor{blue} \color{red}

\usepackage{graphicx}
\usepackage{dcolumn}
\usepackage{color}
\usepackage{graphicx}
\usepackage{dcolumn}
\usepackage{bm}
\usepackage{color}
\usepackage{bm}
\usepackage{tikz}
\usetikzlibrary{decorations.pathmorphing,decorations.markings,trees}
\usepackage[utf8]{inputenc}
\usepackage[T1]{fontenc}
\usetikzlibrary{shapes,arrows,positioning,automata,backgrounds,calc,er,patterns}
\tikzstyle{vertex}=[circle, draw, inner sep=0pt, minimum size=6pt]

\usepackage{graphicx}
\usepackage{dcolumn}
\usepackage{bm}
\usepackage{amsmath}
\usepackage{slashed}
\usepackage[utf8]{inputenc}
\usepackage[T1]{fontenc}
\usepackage{ulem}
\usepackage{amsmath}
\usepackage{amssymb}
\usepackage{float}

\usepackage{cancel}

\def\bb    #1{\hbox{\boldmath${#1}$}}
\def\bb    #1{\hbox{\boldmath${#1}$}}

\def\2d{{{}_{\rm 2D}}}         % put it as {\2d} when it is used
\def\4d{{{}_{\rm 4D}}}         % put it as {\4d} when it is used

\begin{document}

\maketitle

\begin{abstract} 
The relativistic two body problem is considered in terms of the action integral  in the case of   two interacting  spinless particles and spin-$1/2$ fermions, interacting by means of  vector and scalar fields.  The Lagrangians governing the dynamics of such   particles are derived in the momentum representation. In the Breit frame, when the two-body system is at rest as whole,  the obtained Lagrangians generate the Klein-Gordon-like and Dirac-like equations  in the (3+1) phase space.  The derived equations  are examined by studying a  para-positronium  in a strong magnetic field. It is shown that  a strong magnetic fields result in essential decreasing the para-positronium life-time.
\end{abstract}

\section{Introduction}

The  relativity  of  time is a crucial obstacle to formulate the relativistic two-body problem in the explicitly covariant form in the coordinate representation, as compared with the non-relativistic case.  The first attempt to describe the relativistic two-body system was made by summing two free Dirac equations,  adding to them interaction potential \cite{Bre'29,Bre'30}, that resulted in the relativistic non-invariant Breit equation.
On other hand,  the manifestly covariant    Bethe–Salpeter equation\cite{BSE} suitable for solving such kind   problems,  was found to have\cite{Nak}  negative norm solutions in the case  of  two one half spin particles. The lack of the approaches mentioned above can  be eliminated with Dirac's Hamiltonian constraint technique\cite{Dir,Ber}, decreasing a number of the Hamiltonian variables by means of establishing relations between them.

Beginning from 80th,  the constraint dynamic ideas\cite{Dir,Ber}  have been  actively and successfully developed in studying the quantum two-body systems,  and applied for describing real compound systems in QED and QCD \cite{Droz,Kal,Teo76,Teo78,Kom,Sad86-1,Sad86-2,Sad89,Cra84,Cra87,Cra90}. Following the constraint dynamics approach the dynamics of two interacting particle systems involving spin-0 bosons and/or spin-(1/2) fermions is studied \cite{Sad86-1},  and is applied for the description of quarkonium state \cite{Sad86-2}. In the framework of the developed method the $4\times 4$ matrix wave function in terms of one of the $2\times 2$ components, to a single equation of the Pauli–Schrödinger type is constructed\cite{Sad94},  the proposed approach itself  is  generalized to the case of N interacting particle\cite{Sad89}. 
Along  with researches\cite{Sad86-1,Sad86-2,Sad89},  by developing previous results\cite{Teo76,Teo78},  the quantum consideration of two spinless\cite{Cra84} and spin-$1/2$\cite{Cra87} particles under  vector and scalar  interaction , have been done, using the  constraint mechanics for two-body Klein-Gordon  and Dirac  equations.   Two-body states in the  case of  QED interaction\cite{Bar87} and when the interaction is realized due to  a  neutral boson\cite{Bar91}, has been studied in terms of the action integral in the coordinate representation  that leads to the non-covariant Breit-like equation in the coordinate space-time.

 In the present paper the two-body problem is studied in terms of the principle of least action.
 Considering the action integral in the momentum representation, we rewrite the action integral in the variables of total  and relative momentum of the two-body system consisting of  spinless and spin-$1/2$ particles in the momentum space.  It is found that such a written  action integral can  be expressed in terms of the effective energy and effective mass of two  interacting particles.  Provided that the relativistic third Newton law takes place the derived action integral turns out to lead to the covariant equation in ${\cal R}^{3+1}$ and in ${\cal R}^{3+1}\bigcup {\cal S}^2\{{\cal R}^3\}$\cite{Teo76} phase spaces for scalar and $1/2$-spin particles, respectively. In the Breit-like frame, when the center mass of two interacting particles  at rest, whereas   the ``relative `` energy of  these  particles are zero,  the motion equations are radically simplified and become the Klein-Gordon and Dirac types equation. The  compound two fermion system in the singlet state in a strong uniform magnetic field is studied in detail.

\section{Two interacting spinless particles }

We consider two scalar particle with   masses $m_1$ and $m_2$ which interact  due to  vector $A_{}^\nu (x)$ and  scalar $S_{}(x)$ fields. The action integral such a system of particles enumerated by subscribes ``1'' and ``2'',  which  consists of two terms is given by a formula

\begin{eqnarray}\label{eq1}
&&{\frak A} =\frac{\lambda_1}{2} \int dx\Big(  \phi^\ast_1 (x) (-i{\overleftarrow \partial}^\nu +  A_{1}^\nu (x)) ( i\partial_\nu + A_{1\nu} (x)) \phi_1 (x)  - \phi^\ast_1 (x) ( m_1 + S_1(x))^2    \phi_1 (x) \Big) +\nonumber \\
&&\frac{\lambda_2}{2} \int dx\Big(  \phi^\ast_2 (x) (-i{\overleftarrow \partial}^\nu +  A_{2}^\nu (x)) ( i\partial_\nu + A_{2\nu} (x)) \phi_2 (x)  - \phi^\ast_2 (x)( m_2 + S_2(x))^2    \phi_2 (x) \Big), 
\end{eqnarray}
where $ \phi_{1,2} (x) $ denotes the scalar fields, $A_{1,2}^\nu (x)$ and $S_{1,2} (x)$ are the vector and scalar fields, $\lambda_{1,2}$ are the Lagrange multipliers. The coupling constants are included in $A_{1,2}^\nu (x)$ and $S_{1,2} (x)$.

Further consideration demands  to discuss the structure of the interaction terms in Eq.(\ref{eq1}). The field $A_{(1,2) }^\nu (x)$ acting on a particle "1" is obviously governed by  the current generated by a  particle  "2" $j_{(2,1) }^\nu (x)$, and contrariwise. Therefore, we have for  $A_{(1,2) }^\nu (x)$

\begin{eqnarray}\label{eq2}
&& A_{1,2}^\nu (x)= \int  {\cal D}^{\nu \mu}_{1,2} (x-y) (j_{2,1( int)})_\mu (y)  dy , 
\end{eqnarray}
where $ {\cal D}^{\nu \mu}_{1,2} (x-y)$ is the exact propagator of the  vector field created by the second particle and acting on the first one. We will think below that such an interacting field does not exist as a free. Going into the momentum representation according to formulas

\begin{eqnarray}\label{eq2-1}
&& f^\nu (x)= \int \frac{dk}{(2\pi)^4} f^\nu (k) \exp {(-ikx)} , ~~f^\nu (k)= \int {dx} f^\nu (x) \exp {(ikx)}
\end{eqnarray}

 we derive from Eq.(\ref{eq2})

\begin{eqnarray}\label{eq3}
&& A_{1,2}^\nu (x)= \int \frac{dk}{(2\pi)^4} {\cal D}^{\nu \mu}_{1,2} (k) (j_{2,1})_\mu (k) \exp {(-ikx)} .
\end{eqnarray}

Substitution of Eq.(\ref{eq3}) into the action integral Eq.(\ref{eq1}) , in general, leads   to  the unclosed  set  of the  coupled   equations. Therefore,  provided that ${\cal D}_{1,2} (k) j_{2,1}^\nu (k)$ is known or can be perturbatively calculated\cite{Log}, we introduce

 \begin{subequations}\label{eq4}
\begin{eqnarray}
&& A_{1,2 }^\nu (x)\equiv {\cal A}_{1,2 }^\nu (x)= \int \frac{dk}{(2\pi)^4} {\cal A}_{1,2 }^\nu (k) \exp {(-ikx)}   \\
&&{\cal A}_{1,2 }^\nu (k)=  {\cal D}^{\nu \mu}_{1,2} (k) (j_{2,1})_\mu (k) .
\end{eqnarray}
\end{subequations}

Following the same way we can write

\begin{eqnarray}\label{eq5}
&& S_{1,2 } (x)\equiv {\cal S}_{1,2 } (x)= \int \frac{dk}{(2\pi)^4} {\cal S}_{1,2} (k) \exp {(-ikx)}   
\end{eqnarray}
 $ A_{1,2 }^\nu (x)$ and  $S_{1,2} (x) $ introduced  by such a way mean going from real potentials to some quasi-potentials that was done everywhere in Ref.\cite{Sad86-1,Sad86-2,Sad89,Cra84,Cra87,Cra90,Cra01,Cra03,Cra10} earlier. 

We set up  that the fields $\phi_{1,2}  (x)$ are normalized as follows

\begin{eqnarray}\label{eq6}
&& \int dx \phi^\ast_{1,2} (x)\phi_{1,2} (x)  = C_{1,2}< +\infty , 
\end{eqnarray}
where the  value of the constant $C_{1,2}$ is not important as it will be seen below.
Multiplying the first term in Eq.(\ref{eq1}) by $C_1$ but the second term by  $C_2$ and taking $\lambda_{1,2}$ so that $\lambda_{1,2} C_{1,2}=1$, we get
 
\begin{eqnarray}\label{eq7}
&&{\frak A} =\frac{1}{2} \int dx_1 dx_2\Bigg\{\Big(  \phi^\ast_1 (x_1)  \phi_2^\ast ( x_2)(-i{\overleftarrow \partial}^\nu (1) +  {\cal A}_{1}^\nu (x_1))) ( i\partial_\nu (1) +  {\cal A}_{1\nu} (x_1)) \phi_1 (x_1)  \phi_2 ( x_2)-\nonumber \\
&& \phi^\ast_1 (x_1)  \phi_2^\ast ( x_2) (m_1 +{\cal S}_1 (x_1))^2 \phi_1 (x_1)  \phi_2 ( x_2)\Big) +\nonumber \\
&&   \Big(  \phi^\ast_1 (x_1)  \phi^\ast_2 ( x_2)(-i{\overleftarrow \partial}^\nu (2)+ {\cal A}_{2}^\nu (x_2))) ( i\partial_\nu (2) + {\cal A}_{2\nu} (x_2)) \phi_1 (x_1)  \phi_2 ( x_2) - \nonumber \\
&&\phi^\ast_1 (x_1)  \phi_2^\ast ( x_2) (m_2 +{\cal S}_2(x_2))^2  \phi_1 (x_1)  \phi_2 ( x_2)\Big) \Bigg\}, 
\end{eqnarray}

 Going from  $\phi_{1} (x_1)  \phi_{2} ( x_2)$ to $\phi_1 (p_1)  \phi_2 ( p_2)$  in Eq.(\ref{eq1}), 
 
 \begin{eqnarray}\label{eq9}
&&\phi_{1} (x_1)  \phi_{2} ( x_2)=\int \frac{dx_1 dx_2}{(2\pi)^8} e^{-ix_1 p_1 -ix_2 p_2 }\phi_1 (p_1)  \phi_2 ( p_2), 
\end{eqnarray}
and taking into account of Eqs.(\ref{eq4}), (\ref{eq5}), we obtain  the action integral in the momentum representation in the case when $A_{1,2} (x) = A_{1,2} (x_1-x_2)$

\begin{eqnarray}\label{eq10}
{\frak A}& =&  \frac{1}{2}\int \frac{dp_1 dp_2 dq d q'}{(2\pi)^{16}} \phi^\ast_1 (p_1+q+q')\phi_2^\ast ( p_2-q-q')\Big( (2\pi)^{8} \delta (q) \delta (q') (p_1^2-m_1^2)\nonumber \\
&&+2 (2\pi)^{4}  ( p_1 {\cal A}_{1} (q )+\frac{1}{2}q {\cal A}_{1} (q )-m_1 {\cal S}_1 (q))\delta (q')+({\cal A}_{1\nu}  (q)  {\cal A}_{1} ^\nu (q')) -{\cal S}_{1  } (q) {\cal S}_{1  } (q'))\Big)\phi_1 (p_1) \phi_2 (p_2) 
\nonumber \\
&& + \frac{1}{2}\int \frac{dp_1 dp_2 dq d q'}{(2\pi)^{16}} \phi^\ast_1 (p_1+q+q')\phi_2^\ast ( p_2-q-q')\Big( (2\pi)^{8} \delta (q) \delta (q') (p_2^2-m_2^2)\nonumber \\
&&+2 (2\pi)^{4}  ( p_2 {\cal A}_{2} (q )+\frac{1}{2}q {\cal A}_{2} (q )-m_2 {\cal S}_2 (q))\delta (q')+({\cal A}_{2\nu}  (q)  {\cal A}_{2} ^\nu (q')) -{\cal S}_{2  } (q) {\cal S}_{2 } (q'))\Big)\phi_1 (p_1) \phi_2 (p_2) \nonumber \\
\end{eqnarray}

We introduce new variables\cite{Teo71,Teo76}

 \begin{subequations}\label{eq11}
\begin{eqnarray}
&&~P=p_1+p_2, ~~~p=\mu_1 p_2 - \mu_2 p_1,~~~p_1 = \mu_1 P - p, ~~~p_2 = \mu_2 P + p,~~~~~~~~ \\
&&\mu_1 = \frac{1}{2}\left(1+ \frac{m_1^2 -m_2^2 }{M^2}\right), ~~\mu_2 =\frac{1}{2}\left(1- \frac{m_1^2 -m_2^2 }{M^2}\right), \\
&&\mu_1 +\mu_2 = 1, ~~ ~~ M= |p_1+p_2 | = |P|
\end{eqnarray}
\end{subequations}
where  $P$ is the total momentum of two particles,  $M=|P|=|p_1+p_2|$ is the invariant mass of two particles. 
Under    $(p,P)$  variables Eq.(\ref{eq10}) the action integral (\ref{eq2}) is written as

\begin{eqnarray}\label{eq14}
&&{\frak A} =  \int  dq d q' \frac{dP dp}{(2\pi)^{8}} \phi^\ast (\mu_1 P - p|\mu_2 P + p) \Bigg\{\Bigg[\Big( -\bb p^2+  E_w^2 - m_w^2- Pp\frac{m_1^2 -m_2^2 }{M^2}\Big)  \delta(q) \delta(q')\nonumber \\
&&+\frac{1}{2} \Big( 2 (2\pi)^{-4}(  (\mu_1 P - p+\frac{1}{2}q) {\cal A}_{1}(q )-m_1 {\cal S}_1 (q))\delta (q')\nonumber \\
&&+(({\cal A}_{1\nu}  (q)  {\cal A}_{1} ^\nu (q')) -{\cal S}_{1  } (q) {\cal S}_{1  } (q'))\Big)\Bigg] \phi (\mu_1 P - p+q+q'|\mu_2 P + p-q-q')  +\nonumber \\
&&+\Bigg[\frac{1}{2} \Big( 2 (2\pi)^{-4} ) ((\mu_2 P + p+\frac{1}{2}q) {\cal A}_{2}(q )-m_2 {\cal S}_2 (q))\delta (q')\nonumber \\
&&+(2\pi)^{-8}(({\cal A}_{2\nu}  (q)  {\cal A}_{2} ^\nu (q')) -{\cal S}_{2  } (q) {\cal S}_{2  } (q'))\Big) \Bigg] \phi (\mu_1 P - p+q+q'|\mu_2 P + p-q-q')  \Bigg\}
\end{eqnarray}
where  we introduce the two-body wave function
\begin{eqnarray}\label{eq15}
&&\phi  ( p_1,p_2 ) = \phi_1  (p_1 )\phi_2 (p_2 )=\phi (\mu_1 P - p| \mu_2 P + p),
\end{eqnarray}
and an effective energy and an effective mass of this two-body  field

\begin{eqnarray}
&&E_w= \frac{1}{2M} (M^2- (m_1^2 +m_2^2 ) ), ~~~m_w =\frac{ m_1 m_2   }{M}.
\end{eqnarray}

Following\cite{Teo76,Teo78,Cra84,Cra87} we demand

\begin{eqnarray}\label{eq12}
&&Pp =0 ,
\end{eqnarray}
that means  in the quantum case

\begin{eqnarray}\label{eq13}
&&(Pp)\phi  ( p,P )= (pP)\phi  ( p,P ) =0 ,
\end{eqnarray}
where $\phi (p,P) $ is a two-particle wave function.
Eqs.(\ref{eq12}),  (\ref{eq13}) are   the relativistic version of the third Newton law\cite{Cra84,Cra87}.  

Let us  go to the so called Breit frame, where $\bb P= \bb p_1 + \bb p_2 =0$  and $p^0 = 0$. In such a frame Eqs.(\ref{eq12}), (\ref{eq13}) takes place automatically . As a result, we get in the Breit frame

\begin{eqnarray}\label{eq16}
&&{\frak A} =  \int  dq d q' \frac{dP dp}{(2\pi)^{8}} \varphi^\ast (P^0 ,  - \bb p | P^0 + \bb p) \Bigg\{\Bigg[\Big( - \bb p^2+  E_w^2 - m_w^2\Big)  \delta(q) \delta(q')\nonumber \\
&&+\frac{1}{2} \Big( 2 (2\pi)^{-4}(  (\mu_1 P - p+\frac{1}{2}q) {\cal A}_{1}(q )-m_1 {\cal S}_1 (q))\delta (q')\nonumber \\
&&+(({\cal A}_{1\nu}  (q)  {\cal A}_{1} ^\nu (q')) -{\cal S}_{1  } (q) {\cal S}_{1  } (q'))\Big)\Bigg] \varphi ( P^0 +q^0+q'^0,  - \bb p+\bb q+ \bb q' |  P^0-q^0-q'^0 , \bb p-\bb q- \bb q')  +\nonumber \\
&&+\Bigg[\frac{1}{2} \Big( 2 (2\pi)^{-4}(  (\mu_2 P + p+\frac{1}{2}q) {\cal A}_{2}(q )-m_2 {\cal S}_2 (q))\delta (q'+(2\pi)^{-8}(({\cal A}_{2\nu}  (q)  {\cal A}_{2} ^\nu (q')) -{\cal S}_{2  } (q) {\cal S}_{2  } (q'))\Big) \Bigg] \nonumber \\
&&\times \varphi ( P^0 +q^0+q'^0,  - \bb p+\bb q+ \bb q' |  P^0-q^0-q'^0 , \bb p-\bb q- \bb q') \Bigg\}, 
\end{eqnarray}
where we establish

\begin{eqnarray}\label{eq16}
\varphi^\ast (P^0 ,  - \bb p|P^0 + \bb p) \varphi (P^0 ,  - \bb p|P^0 + \bb p) =  \phi^\ast (\mu_1 P - p|\mu_2 P + p) \phi (\mu_1 P - p| \mu_2 P + p) \delta (\bb P) \delta( p^0 ) 
\end{eqnarray}
Note, that  $P=(P^0, 0) , p=(0,\bb p)$  in latest  equation because the $\delta$-function.
The  introduced  function $\varphi (P^0 ,  - \bb p|P^0 + \bb p)$  describes the  field of a two particle system, which  is averaged with respect to all possible relative times and with respect to all possible positions of a center mass.
 
 We note that changing a sign before $\bb p$ means changing a particle number.  
 Since  the variables of the first particle are before the vertical line in the wave function given by Eq.(\ref{eq15}), whereas   the variables  are of  the second particle   after it, we derive

\begin{eqnarray}\label{eq17}
&&{\frak A} =  \int  dq d q' \frac{dP^0 d\bb p}{(2\pi)^{8}} \varphi^\ast ( P^0 ,\bb p) \Bigg\{\Big(-\bb  p^2+  E_w^2 - m_w^2\Big)  \delta(q) \delta(q')\nonumber \\
&&+  (2\pi)^{-4}  \Big((\mu_1 P - p+\frac{1}{2}q) {\cal A}_{1}(q )-m_1 {\cal S}_1 (q)+(\mu_2 P + p+\frac{1}{2}q) {\cal A}_{2}(q )-m_2 {\cal S}_2 (q)\Big)\delta (q')\nonumber \\
&&+\frac{1}{2}(2\pi)^{-8}\Big({\cal A}_{1\nu}  (q)  {\cal A}_{1} ^\nu (q') -{\cal S}_{1  } (q) {\cal S}_{1  } (q')+{\cal A}_{2\nu}  (q)  {\cal A}_{2} ^\nu (q') -{\cal S}_{2  } (q) {\cal S}_{2  } (q')\Big)\Bigg\}\varphi (   P^0+q^0+q'^0,  \bb  p+\bb q+\bb q')  \nonumber \\
\end{eqnarray}

The demand of carrying out  the 3rd Newton law in 3D space for   interacting particles  is  achieved by introducing  new vector and scalar fields according to formulas  
   
\begin{subequations}\label{eq18}
\begin{eqnarray}
&&{\cal  A}_1 =2\pi~ \Big({\cal A}^0\frac{2E_w\big(\mu_1 -\mu_2\sqrt{(P^0)^2 (\mu_1^2+\mu_2^2)/2E^2_w -1}\big)}{P^0(\mu_1^2+\mu_2^2)}, - \bb {\cal A}\Big)\delta ( q^0 )\\
&&{\cal  A}_2 = 2\pi~\Big({\cal A}^0\frac{2E_w\big(\mu_2 +\mu_1\sqrt{(P^0)^2 (\mu_1^2+\mu_2^2)/2E^2_w -1}\big)}{P^0(\mu_1^2+\mu_2^2)},  \bb {\cal A}\Big)\delta ( q^0 )
\end{eqnarray}
\end{subequations}

\begin{subequations}\label{eq19}
\begin{eqnarray}
&& {\cal S}_{1}   = 2\pi a~{\cal S } \delta ( q^0 ) \\
&& {\cal S}_{2 } =2\pi b ~{\cal S }  \delta ( q^0 ),  \\
&&a=\frac{m_2\sqrt2 +2m_w}{m_1+m_2}, ~~~~b=\frac{m_1\sqrt2 -2m_w}{m_1+m_2}.
\end{eqnarray}
\end{subequations}
 The factors,  multiplying $A^0 $,  and which modernize the scalar field, have be taken to keep the gauge invariance of a new Lagrangian.

As a result, we derive

\begin{eqnarray}\label{eq20}
&&{\frak A} =  \int  d\bb q d \bb q' \frac{dP^0 d\bb p}{(2\pi)^{8}} \varphi^\ast (P^0 ,  \bb p) \Bigg\{\Big( -\bb p^2+  E_w^2 - m_w^2\Big)  \delta(\bb q) \delta(\bb q')\nonumber \\
&&+  (2\pi)^{-3}  \Big((2 E_w {\cal A}^0 (\bb q) - 2\bb p \bb {\cal A} (\bb q)- \bb q  \bb {\cal A} (\bb q))- 2 m_w  {\cal S}(\bb q)\Big)\delta (\bb q')\nonumber \\&&+(2\pi)^{-6}\Big( {\cal A}^0 (\bb q)  {\cal A}^0 (\bb q')- \bb {\cal A}(\bb q)\cdot   \bb {\cal A}(\bb q') -  {\cal S}(\bb q) {\cal S}(\bb q')\Big)\Bigg\}\varphi (  P^0,  \bb  p-\bb q-\bb q')  \nonumber \\
\end{eqnarray}
The  Lagrangian  in the action integral (\ref{eq20}) is independent explicitly on $P^0$ that  means the conservation of $P^0$, and leads to independence of the wave function on $P^0$ as a variable. Then, 
varying the derived action integral we go to the motion equation in the momentum representation

\begin{eqnarray}\label{eq21}
&&\Big( -\bb p^2+  E_w^2 - m_w^2\Big)\varphi (  \bb  p)  +   \int  \frac{d\bb q }{  (2\pi)^{3}}  \Big((2 E_w {\cal A}^0 (\bb q) - 2\bb p \bb {\cal A} (\bb q)- \bb q  \bb {\cal A} (\bb q))+ 2 m_w  {\cal S}(\bb q)\Big)\varphi (  \bb  p-\bb q)\nonumber \\&&+ \int \frac{ d\bb q d \bb q'}{ (2\pi)^{6}}\Big( {\cal A}^0 (\bb q)  {\cal A}^0 (\bb q')- \bb {\cal A}(\bb q)\cdot   \bb {\cal A}(\bb q') +  {\cal S}(\bb q) {\cal S}(\bb q')\Big)\varphi (   \bb  p-\bb q-\bb q') =0 \nonumber \\
\end{eqnarray}

Coming back  to the coordinate representation we obtain the motion equation in the Schrödinger  form

\begin{eqnarray}\label{eq22}
&&\Big(- (\bb p+\bb{\cal A} )^2+ ( E_w+ {\cal A}^0)^2 - (m_w+ {\cal S})^2\Big)  \varphi ( \bb r) =0,
\end{eqnarray}
where $\bb p=-i\nabla$ is the relative momentum of the two-body system.
The derived equation is gauge invariant , and can be reduced to Eq.(50) obtained earlier\cite{Cra87}, shifting   a vector field due to the gauge invariance of Eq.(\ref{eq21}). 

We note that the equation (\ref{eq22}) is invariant with respect to rotations in the ${\cal R}^{3+1} $ phase space.

\section{Two interacting  spin-$\frac{1}{2}$ particles }

We consider   two spin one half  particles with   masses $m_1$ and $m_2$ which interact each other  by means of   vector  and  scalar  fields , which are $A_{int}^\nu (x)$ and $S_{int}(x)$, respectively, We also take that these particles are in external vector $A_{ext}^\nu (x)$ and  scalar $S_{ext}(x)$ fields. The action integral such a system of particles which  consists of two terms is

\begin{eqnarray}\label{eq23}
&&{\frak A}=\lambda_1 \int dx_1 {\bar \Psi }_1 (x_1)\Big( i\gamma^\nu\partial_\nu + \gamma^\nu A_{1\nu} (x_1)-(m_1+ S_1 (x_1 )\Big) { \Psi }_1 (x_1) + \nonumber \\
&&\lambda_2 \int dx_2 {\bar \Psi }_2 (x_2)\Big( i\gamma^\nu\partial_\nu + \gamma^\nu A_{2 \nu} (x_2)-(m_2+S_2 (x_2) )\Big) { \Psi }_2 (x_2), 
\end{eqnarray}
where the sense of all symbols introduced in Eq.(\ref{eq22}) is  literally the same as it has been  in Eqs.(\ref{eq2})- (\ref{eq5}) of  the previous section. We select  the Lagrange multipliers $\lambda_1$ and $\lambda_2$  to satisfy   relations

\begin{eqnarray}\label{eq24}
&& \frac{\lambda_{1}}{(2 \int dx_1 {\bar \Psi }_1 (x_1)(m_1 + S(x_1)) { \Psi }_1(x_1))( \int dx_2 {\bar \Psi }_2 (x_2) { \Psi }_2(x_2))} =1, \nonumber \\
&& \frac{\lambda_{2}}{(2 \int dx_2 {\bar \Psi }_2 (x_2)(m_2 + S(x_2)) { \Psi }_2(x_2))( \int dx_1 {\bar \Psi }_1 (x_1) { \Psi }_1(x_1))} =1.
\end{eqnarray}
Then, using the motion equation for a single particle, we get the following  for the action integral (\ref{eq23})

\begin{eqnarray}\label{eq25}
&&{\frak A}= \int dx_1   dx_2  {\bar \Psi }_1 (x_1) {\bar \Psi }_2 (x_2)\Big( \gamma^\nu(i\partial_\nu (1)+  A_{1\nu} (x_1))+(m_1+ S_1 (x_1 )\Big)\nonumber \\
&&\times \Big( \gamma^\nu(i\partial_\nu (1)+  A_{1\nu} (x_1))-(m_1+ S_1 (x_1 )\Big) { \Psi }_1 (x_1) { \Psi }_2 (x_2) + \nonumber \\
&&\int dx_1   dx_2  {\bar \Psi }_1 (x_1) {\bar \Psi }_2 (x_2)\Big( \gamma^\nu(i\partial_\nu(2) +  A_{2\nu} (x_2))+(m_2+ S_ (x_2 )\Big)\nonumber \\
&&\times \Big( \gamma^\nu(i\partial_\nu (2)+  A_{2\nu} (x_1))-(m_2+ S_2 (x_2 )\Big) { \Psi }_1 (x_1) { \Psi }_2 (x_2) 
\end{eqnarray}
The latest equation can be written as

\begin{eqnarray}\label{eq26}
&&{\frak A} = \pm \int dx_1 dx_2\Bigg\{\Big(   \phi^\ast (x_1, x_2)(-i{\overleftarrow \partial}^\nu (1) +  {\cal A}_{1}^\nu (x_1))) ( i\partial_\nu (1) +  {\cal A}_{1\nu} (x_1))  \phi (x_1, x_2)-\nonumber \\
&& \phi^\ast (x_1, x_2) (m_1 +{\cal S}_1 (x_1))^2    \phi (x_1, x_2)\Big) +\nonumber \\
&&   \Big(  \phi^\ast (x_1, x_2)(-i{\overleftarrow \partial}^\nu (2)+ {\cal A}_{2}^\nu (x_2))) ( i\partial_\nu (2) + {\cal A}_{2\nu} (x_2))  \phi (x_1, x_2) - \nonumber \\
&&\phi^\ast (x_1, x_2) (m_2 +{\cal S}_2(x_2))^2  \phi (x_1, x_2)\Big) \Bigg\}, 
\end{eqnarray}
where $ \phi (x_1, x_2) \equiv \phi_1 (x_1)  \phi_2 ( x_2)$ is one of the 16 components of the direct product of two Dirac bispinors ${ \Psi }_1 (x_1) \otimes { \Psi }_2 (x_2)$. The signs plus or minus which have  arisen before the integral come from  the direct production of two $\gamma^0$-matrices.  They obviously do not affect the motion equation for each of the components  $ \phi (x_1, x_2)$.

The action integral (\ref{eq26}) coincides with ${\frak A}$ given by (\ref{eq7}) up to notations. Therefore,  we make the same steps, which have been done, starting from Eq.(\ref{eq8}) and finishing by  Eq.(\ref{eq22}). As a result, provided that Eqs.(\ref{eq12}), (\ref{eq13}) is satisfied, we obtain in the Breit frame ($\bb P= \bb p_1 + \bb p_2 =0, ~ p^0 = 0$) 

\begin{eqnarray}\label{eq27}
&&\Big(- ( {\bb p}+\bb{\cal A} )^2+ ( E_w + {\cal A}^0)^2 - (m_w+ {\cal S})^2\Big)  \varphi_i  ( \bb r) =0,
\end{eqnarray}
where $\varphi_i  ( \bb r)$ is a  components of the direct product of two Dirac bispinors ${ \Psi }_1 (x_1) \otimes { \Psi }_2 (x_2)$.

To  derive the motion equation Eq.(\ref{eq27})  in the covariant  form  we follow  P.A.M.Dirac, and 
 rewrite it as

\begin{eqnarray}\label{eq29}
&&[( p_\nu+ {\cal A}_\nu) g^{\nu \mu 
}( p_\mu+ {\cal A}_\mu )  -(m_w+ {\cal S})^2  ]\varphi_i  ( \bb r) =0.
\end{eqnarray}
To obtain the motion equation in the needed form   we express the metric tensor in terms of the $\gamma$-matrices of  the $(16 \times 16)$ dimension.

Eq.(\ref{eq11}) which  introduces new variables  implies going to a new basis in the $\gamma$-matrix space so that new and old  $\gamma$-matrices are related to each other by means of equations

\begin{eqnarray}\label{eq28}
&&{\hat \Gamma} =\mu_1  \stackrel {(1)} {\gamma}\otimes \stackrel {(2)} {I}+\mu_2  \stackrel {(2)} {\gamma}\otimes \stackrel {(1)} {I},\nonumber \\
&& {\hat{  \gamma}}=  \stackrel {(2)} {{ \gamma}}\otimes \stackrel {(1)} {I}-  \stackrel {(1)} { \gamma}\otimes \stackrel {(2)} {I}
\end{eqnarray}
where the hatted $\hat\gamma$-matrices  form a new basis, the number  in brackets which over a symbol  means   that this matrix acts on the variables of the corresponding particle, $\stackrel {(1,2)} {I}$ is the unit matrix in the space of the first or second  particle, respectively. In the Breit frame of reference,  when $\bb P= \bb p_1 + \bb p_2 =0, ~~ p^0 = 0$,  this new basis consists of matrices $({\hat \Gamma}^0 , {\hat{ \bb{ \gamma}}})$

\begin{eqnarray}\label{eq30}
&&{\hat \Gamma}^0 =\mu_1  \stackrel {(1)} {\gamma^0}\otimes \stackrel {(2)} {I}+\mu_2  \stackrel {(2)} {\gamma^0}\otimes \stackrel {(1)} {I},\nonumber \\
&& {\hat{ \bb{ \gamma}}}=  \stackrel {(2)} {{\bb \gamma}}\otimes \stackrel {(1)} {I}-  \stackrel {(1)} {\bb \gamma}\otimes \stackrel {(2)} {I}
\end{eqnarray}

These matrices satisfy the following transformation relations

\begin{subequations}\label{eq31}
\begin{eqnarray}
&&{\hat \Gamma}^0{\hat \Gamma}^0 +{\hat \Gamma}^0{\hat \Gamma}^0= 2 \stackrel {(1+2)} {I};\\
&&{\hat \Gamma}^0 {\hat{  \gamma^a}}+ {\hat{  \gamma^a}}{\hat \Gamma}^0=0, ~~a=1,2,3;\\
&&{\hat{ \gamma^a}}{\hat{ \gamma^b}} +{\hat{ \gamma^b}}{\hat{ \gamma^a}}=-4 \delta^{ab} \stackrel {(1+2)} {I}+4 (s_1^a s_2^b+s_2^b s_1^a+ s_2^a s_1^b+s_1^b s_2^a) \stackrel {(1+2)} {I}, 
\end{eqnarray}
\end{subequations}
where  $\bb s_1$ and $\bb s_2$ are particle spins, $\delta^{ab}$ is the Kronecker symbol, $\stackrel {(1+2)} {I}$  is the $16\times 16$ unit matrix. Such transformation relations allow us to write down the metric tensor in terms of  new  $\gamma$- matrices as follows

\begin{subequations}\label{eq32}
\begin{eqnarray}
&&g^{\nu \mu }=\Gamma^\nu \Gamma^\mu - (s_1^\nu s_2^\mu+s_2^\mu s_1^\nu+ s_2^\nu s_1^\mu+s_1^\mu s_2^\nu) \stackrel {(1+2)} {I}, \\
&&\Gamma^\nu = ({\hat \Gamma}^0, \frac{1}{2}{\hat{ \gamma^a}} )\equiv ({ \Gamma}^0, {{\bb \Gamma}} ) ,
\end{eqnarray}
\end{subequations} 
where $s^0_{1,2} =0 $. 

Substituting Eqs.(\ref{eq32}) into the formulas (\ref{eq27}), (\ref{eq29}) we obtain

\begin{eqnarray}\label{eq33a}
&&\Big( \Gamma^0 ( E_w+ {\cal A}^0)  - \bb \Gamma ({\bb p}+\bb{\cal A} )  +\bb s_1 ( {\bb p}+\bb{\cal A} ) + \bb s_2 ( {\bb p}+\bb{\cal A} )- (m_w+ {\cal S})\Big)  \Psi ( \bb r) =0,
\end{eqnarray}
where $\Psi ( \bb r)$ is the 16 component spinor, the later three terms in Eq.(\ref{eq33a}) is assumed to be multiplied by $\stackrel {(1+2)} {I}$. The obtained equation is invariant with respect to rotations in the  hyperspace consisting of the phase space ${\cal R}^{3+1}$ and two-dimensional sphere ${\cal S}^2\{{\cal R}^{3}\}$ in ${\cal R}^{3}$.
When the two-particle system is in the singlet state   Eq.(\ref{eq33a}) is simplified.  In this case  $\bb s_1 = - \bb s_2$ , $\Gamma^\nu = \gamma^\nu =  ( \gamma^0, \bb \gamma )$ that leads to the standard Dirac equation for a single particle

\begin{eqnarray}\label{eq33}
&&\Big( \gamma^0 ( E_w + {\cal A}^0)  - \bb \gamma ( {\bb p}+\bb{\cal A} ) - (m_w+ {\cal S})\Big)  \Psi ( \bb r) =0,
\end{eqnarray}
where $\Psi ( \bb r)$ is the four component  Dirac bispinor provided that is governs a spinless state. 

\section{Para-positronium in a strong magnetic field}

We illustrate    the motion equations, studying real physical objects. Let us  consider para-positronium in a strong uniform magnetic field. Such a problem arises  in astrophysics. The electromagnetic signals coming from astrophysical object such as pulsars or neutron stars\cite{Bro,Bus},  whose 
energy is about $0.511 MeV$, likely  correspond to the electromagnetic  decay of the ground state para-positrinium in two photons.  In the case when the electrically neutral tow particle system is static external fields Eq.(\ref{eq33a}) keeps to be correct. Then, directing a magnetic field $\bb B$ along $OZ$ axis and going to the second order equitation, we get from Eq.(\ref{eq33}), going by the standard way\cite{LL3,LL4}

\begin{eqnarray}\label{eq34}
&&\Big(\triangle +\frac{2(4 \pi \alpha) E_w }{r}+\frac{(4 \pi \alpha)^2 }{r^2}+\frac{i}{2}\nabla (\bb B\times \bb \rho) -\frac{1}{4} e^2  B^2 \rho^2 \Big)\Psi (\bb r)=( m_w^2 -E_w^2)\Psi (\bb r)
\end{eqnarray}

In obtaining the latest equation we take

\begin{eqnarray}\label{eq35}
&&A^0 = -\frac{4 \pi \alpha}{r}, ~~~{\cal A} = \frac{1}{2}(\bb B\times \bb \rho), 
\end{eqnarray}
where the cylindrical coordinate $\bb r = \bb \rho +\bb e_z z$ is used, $\alpha$ is the fine structure constant.

We study the ground state of  a para-positronium in a strong magnetic field $\bb B$ when the magnetic length $ a = (|e| B)^{-/12}$ is  much smaller   than the positronium Bohr radius $a_B = 2 /m_e e^2$, where  $e$ and $m_e$ are a charge and electron mass, respectively. The wave function governing  the transverse motion of a particle in the ground state in a magnetic field  is\cite{LL3}

\begin{eqnarray}\label{eq36}
&&\psi_{m=0,n=0} (\bb \rho) =\frac{1}{a\sqrt{2 \pi}} e^{-\frac{\rho^2}{4 a^2}}, 
\end{eqnarray}
where $m$  is the projection of an angular momentum along the $\bb B$ direction,  $n$ is a radial quantum number.  We look for the solution of Eq.(\ref{eq34}) in a form

\begin{eqnarray}\label{eq37}
&&\Psi (\bb r) = \psi (z) ~\psi_{m=0,n=0} (\bb \rho). 
\end{eqnarray}
Substituting $\Psi (\bb r)$ given by  Eq.(\ref{eq37}) into Eq.(\ref{eq34}), we derive

\begin{eqnarray}\label{eq38}
&&\Big(\frac{d^2}{d\xi^2} +  2^{1/2}\sqrt{\pi}(4 \pi \alpha) m_w \varepsilon_w a e^{\xi^2/2} erfc (|\xi| /\sqrt 2) +\frac{ (4 \pi \alpha)^2 e^{\xi^2/2}}{2} E_1 (\xi^2/2) \Big)\psi (\xi)=\varepsilon^2\psi (\xi), \nonumber \\
\end{eqnarray}
where we introduce   $erfc (x)$ is the complementary error function, $E_1(x)$ is the integral exponent\cite{Gra}, and

\begin{eqnarray}\label{eq39}
z /a \equiv \xi, ~~~ ( m_w^2 -E_w^2+(1/a^2))a^2 \equiv \varepsilon^2, ~~~~\varepsilon_w = E_w /m_w
\end{eqnarray}

 Since  the magnetic field is strong  a particle  is weakly bound with respect to  a motion along the $OZ$ axis.  Therefore, we look for a  solution of Eq.(\ref{eq38}) in a form

\begin{eqnarray}\label{eq40}
\psi (\xi) =\sqrt{ k} \exp ( - k |\xi|)
\end{eqnarray}
where $k=\sqrt \varepsilon$ is a positive constant.

Substituting such $\psi (\xi)$ into  Eq.(\ref{eq39}) we derive  an equation to obtain $k$.

\begin{eqnarray}\label{eq41}
&&k  =  2^{3/2}\sqrt{\pi}(4 \pi \alpha) m_w \varepsilon_w a \int\limits_0^\infty  d \xi \exp{({\xi^2/2}-2k \xi)}erfc (\xi /\sqrt 2)\nonumber \\
&& +{(4 \pi \alpha)^2}\int\limits_0^\infty  d \xi \exp{({\xi^2/2}-2k \xi)} E_1 (\xi^2/2) .
\end{eqnarray}
The validity limit of the derived equation is $k\ll 1$.  At such $k$ , the second integral \cite{Gra} is approximately  $\pi^{3/2}/\sqrt 2$, whereas the first integral can  be  approximately taken  to be $\ln (a_B/a)/\sqrt \pi$\cite{LL3} after  the dimensional  regularization.  As a result, we find

\begin{eqnarray}\label{eq42}
&& k  =  2^{3/2}(4 \pi \alpha) m_w \varepsilon_w a \ln (a_B/a)+(4 \pi \alpha)^2\frac{\pi^{3/2}}{\sqrt 2}.
\end{eqnarray}

In the non-relativistic  $m_e a \gg\alpha $ case we obtain  $k  =  2^{3/2}(4 \pi \alpha) m_e a\ln (a_B/a)$\cite{Ell},  that leads to

\begin{eqnarray}\label{eq43}
&& E_w =\frac{ m_e}{2} ( 1 + (2 / (m_e a)^2 - 2 (4 \pi \alpha)^2 \ln^2 (a_B/a))
\end{eqnarray}

In the opposite limiting situation $m_e a \ll \alpha $, but provided that  the vacuum polarization effects  in a strong magnetic field are absent\cite{LL4}, we get  $k =(2 \pi)^{7/2} \alpha^2$, that results in

\begin{eqnarray}\label{eq44}
&& E_w =\frac{ 1}{a^2} ( 1 + (m_e a)^4/2 -  (2 \pi)^7  \alpha^4/2)\simeq \frac{ 1}{a^2} ( 1 -  (2 \pi)^7  \alpha^4/2).
\end{eqnarray}
Eqs.(\ref{eq43}), (\ref{eq44}) show the stronger a magnetic field , the less depth of  the para-positronium level  under the Landau zone bottom.

\section{ Conclusion}

We study the relativistic two-body  problem in external fields in terms of the principle of  least action. Starting from  the single particle action integral  for a  spinless particle and for  a fermion with a spin being equal to one half, we derive the Lagrangian governing the dynamics of the two-body system  in external fields beyond the constraint dynamics formalism. The motion equations generated by the derived Lagrangian turn out to be invariant with respect to rotations in the ${\cal R}^{3+1} $ and  ${\cal R}^{3+1}\bigcup  {\cal S}^2\{{\cal R}^{3}\} $ phase spaces for the considered scalar particles and fermions, respectively .  It is shown that the developed approach leads to the same motion equations as in the case of the constraint dynamics what demonstrates a validity of the principle of least action  to  solution of such problems. Based on the derived equations we study the ground para-positronium  state in a strong  uniform magnetic field, and show that increasing a magnetic field results in decreasing the depth  of  the para-positronium level  under the Landau zone bottom.

\end{document}